# Universal Origin of Glassy Relaxation as Recognized by Configuration Pattern-matching

Hai-Bin Yu[1*], Liang Gao[1], Jia-Qi Gao[1] and Konrad Samwer[2*]

[1]Wuhan National High Magnetic Field Center and School of Physics, Huazhong University of Science and Technology, Wuhan 430074, Hubei, China;

[2]I. Physikalisches Institut, Universität Göttingen, D-37077 Göttingen, Germany.

E-mail: haibinyu@hust.edu.cn (HBY), ksamwer@gwdg.de (KS)

**Abstract**

**Relaxation processes are crucial in understanding the structural rearrangements of liquids and amorphous materials. However, the overarching principle that governs these processes across vastly different materials remains an open question. Substantial analysis has been carried out based on the motions of individual particles. Here, alternatively, we propose viewing the global configuration as a single entity. We introduce a global order parameter, namely the inherent structure minimal displacement (IS $D_{min}$), to quantify the variability of configurations by a pattern-matching technique. Through atomic simulations of seven model glass-forming liquids, we unify the influences of temperature, pressure, and perturbation time on the relaxation dissipation, via a scaling law between the mechanical damping factor and IS $D_{min}$. Fundamentally, this scaling reflects the curvature of the local potential energy landscape. Our findings uncover a universal origin of glassy relaxation and offer an alternative approach to studying disordered systems.**

**Introduction**

Glasses are disordered materials that lack the structural long-range order of crystals but behave mechanically like solids. They are typically produced by rapidly cooling a viscous liquid to prevent crystallization. However, the dynamic processes through which liquids acquire amorphous rigidity during cooling remain incompletely understood[1-5].

One hallmark of glasses and glass-forming liquids is their complex relaxation dynamics, which span approximately 12 orders of magnitude in timescales accessible with current technologies. These relaxation processes significantly affect the mechanical and functional properties of glassy materials[6-10]. Understanding how structural rearrangements govern these relaxation processes is crucial for uncovering the nature of glass and designing amorphous materials with improved properties[6-8, 11, 12].

Microscopic explanations of these dynamics are the fundamental components of glass transition research. For instance, we have demonstrated that the β relaxation in metallic glass could arise from string-like cooperative motions[13]. Berthier and co-workers have recently proposed that excess wings result from rearrangements of rare and localized regions over broadly distributed timescales[14]. Guan and colleagues have shown that low-temperature relaxation might be attributed to revisable atomic extrusions with hybrid features of vibration and diffusion motions[15]. Chang and colleagues found evidence that low-temperature relaxation is correlated with the liquid's light temperature properties[16]. More recently, Lukenheimer et al made a correlation of the glass transition temperature and the thermal expansion coefficient with the necessary inclusion of the cooperativity parameter[17], the fragility.

These studies have enhanced our understanding of the nature of glass and glassy motions, providing valuable insights into the design of amorphous materials. However, a general theoretical framework for relaxation dissipation in disordered systems is still lacking. It remains unclear whether there is a unified mechanism for the various relaxation processes observed in diverse glasses[2, 5].

Central to these relaxation measurements using various dynamic spectroscopy techniques is the loss angle δ (also known as the phase-lag angle or damping angle) between the input and output signals, which characterizes the energy dissipation during cyclic perturbation. In particular, δ is widely used as the damping factor of internal friction and has practical engineering applications. It is directly related to density and shear stress fluctuations through the fluctuation-dissipation theorem[18]. Thus, in this work, we focus on a unified understanding of δ.

Developing a general framework that unifies the relaxation dissipation of different glassy systems would be a significant step forward in understanding the nature of glasses and the glass transition. Such a framework could provide a basis for designing new amorphous materials with tailored mechanical and functional properties[19-21].

One approach that may be useful in addressing this issue is the potential energy landscape (PEL)[22-24]. The PEL refers to a potential energy function of an $N$-body system $\Phi(r_1, ..., r_N)$, where the vectors $r_i$ comprise position, orientation, and intermolecular coordinates. It is a multidimensional landscape, for example, in case of $N$ identical atoms, the landscape is a $(3N+1)$-dimensional object.

Essential aspects of the PEL are the presence of local minima, also known as inherent structures (IS). It has been suggested that the way in which a system explores its landscape as a function of temperature can provide insights into its dynamic behavior. However, due to the high dimensionality of the PEL, it is a challenging task to establish quantitative connections between the features of the PEL and material properties[25-30].

In this study, we propose a novel concept called "inherent structure minimal displacement" that utilizes a pattern-matching technique to quantify the variability of IS configurations. We demonstrate that this global order parameter can unify relaxation dissipation into a scaling law, providing a general principle for understanding relaxation in glass-forming liquids.

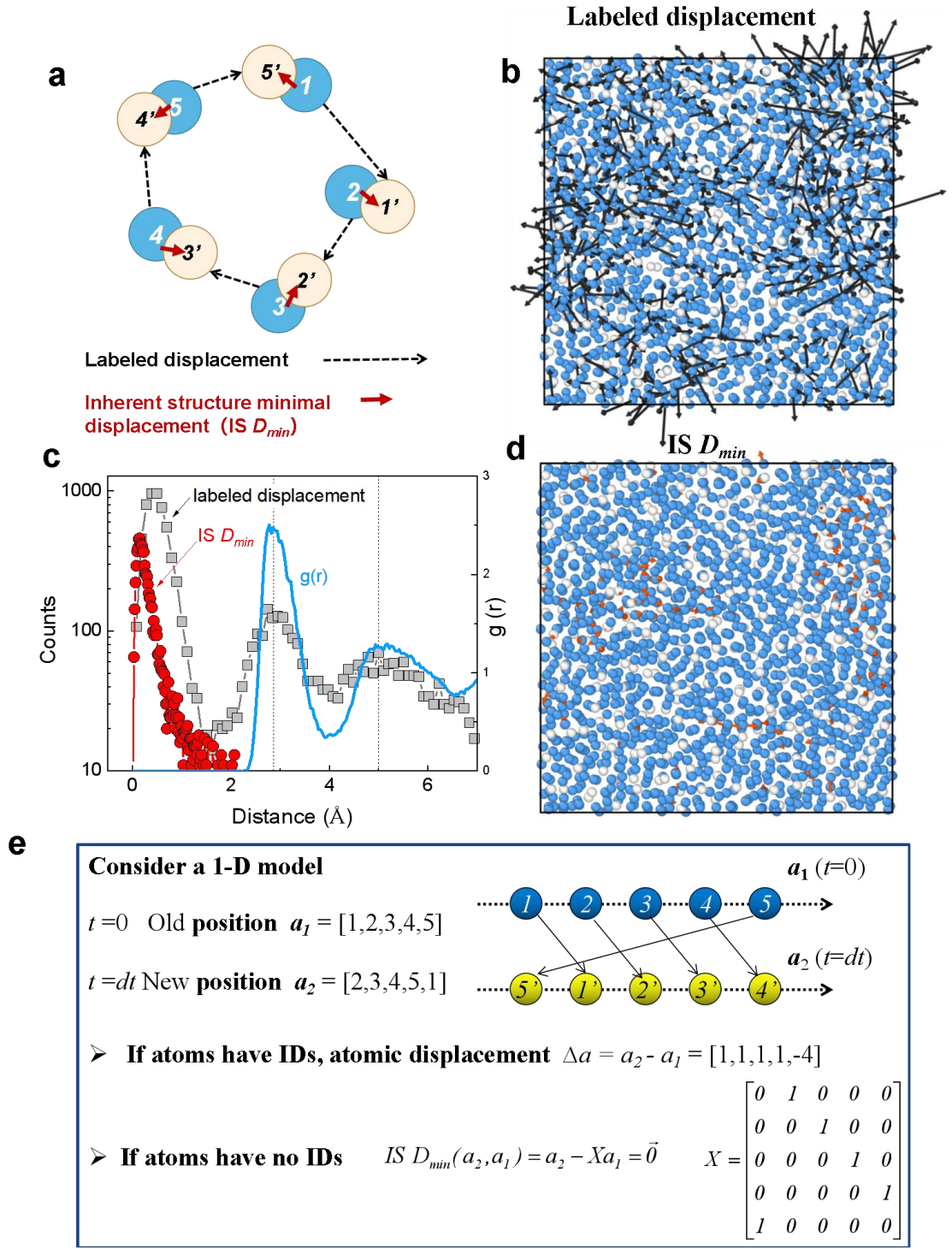

**FIG.1. Definition and quantification of IS $D_{min}$.** This schematic plot in (a) shows the labeled displacements (dashed vector) and the IS $D_{min}$ (solid vector) for each atom. Blue disks with label $i$ (1, 2,.. ) indicate the initial position at $t = t_0$, while yellow disks with label $i'$ (1', 2',.. ) indicate the position at $t = \delta t + t_0$. (b) Typical atomic displacements (black arrows) for an $Al_{85}Sm_{15}$ glass-forming liquid at $T$ = 760 K and $\delta t$

= 2000 ps, at this temperature the main relaxation time $\tau_\alpha$ ($T$ = 760 K) ~ $2.5 \times 10^4$ ps; the plot is sliced from three-dimension model, with the slice thickness 10 Å. The blue and white spheres indicate Al amd Sm atoms respectively. (c) Distribution of displacements and *IS $D_{min}$* (left axis) and the radial distribution function g(r) (right axis). (d) The IS *$D_{min}$* (red arrows) same as condition of (b). (e) a simplified one-dimensional model for considering the compuation of IS *$D_{min}$* and the assocated tansformation matrix *X*.

**Results**

**Definition of IS *$D_{min}$*.** We start by considering an instantaneous configuration at $t=t_0$, as shown schematically in Fig. 1(a), where the atoms are labeled as [*1, 2, 3, 4, 5*] (the blue disks). Later at a time $t = \delta t + t_0$, these atoms move some distances, and a new configuration is arrived as index by [*1', 2', 3', 4', 5'*] (the yellow disks) in Fig.1. The displacement vectors are indicated by dashed lines for each atom.

Now, we consider the inherent structures of the two configurations and consider that all the atoms of the same chemistry are the same (identical particles), in other words, the atoms do not have the labels. Then the minimal displacement between the two configurations [*1, 2, 3 ,4, 5*] and [*1', 2', 3', 4', 5'*] can be represent by the red solid arrows in Fig. 1 (referred as IS *$D_{min}$*, inherent structure minimal displacement).

We note that the *IS $D_{min}$* reflects the minimal cooperative rearrangements, and it can be cast in to a scalar value (see Eqs.2 and 3 in method). It is smaller than the labeled displacement [each dashed vector in Fig.1(a)]. For example, when atomic motions are exactly position replacing and forming a loop-like motion, the IS *$D_{min}$* would be zero, while the labeled displacements would be a finite value.

Since experimental relaxation dynamics are probed by macroscopic tools, we expect that only the variations between two configurations would have detectable effects on relaxation properties. Such a variation can be characterized by the IS *$D_{min}$* as outlined above. This is the major premise and the starting point of the present work. In the following, we show that such a principle indeed works reasonably well for a wide range of glass forming liquids, especially a scaling law will be revealed.

We note that the root-mean-square displacement (RMSD) which is widely used in the atomic dynamic analysis[1], is defined based on the labeled displacements rather than IS $D_{min}$. On the other hand, the definition of IS $D_{min}$ bears some similarity with the overlap parameter suggested by Parisi for spin systems[31], which has also been used in glass research[32]. Their differences are notable. The overlap parameter is a coarse-grained indicator for "similarity" between two configurations. As will be discussed, IS $D_{min}$ characterizes the shortest distance between two neighboring ISs at the same potential level.

The evaluation of IS $D_{min}$ is not as straightforward as RMSD or the overlap parameter, because one has to find an optimum pattern-matching between two configurations for the calculation of IS $D_{min}$. In this work, we use a linear sum assignment algorithm, which is also known as minimum weight matching in bipartite graphs and the "The Hungarian Algorithm" in computer science[33]. Computation details about IS $D_{min}$ are presented at the method section. A simplified one-dimensional (1D) model is given in Fig.1(e), which illustrates that the central idea is to find a transform-matrix that operates on the configuration.

Figure 1(b) and (d) compares the labeled displacements (i.e., associated with RMSD) and the IS $D_{min}$ for a same configuration of an $Al_{85}Sm_{15}$ model glass forming-liquids over a waiting-time about 2000 pico-seconds (ps) in a supercooled state [$\tau_\alpha$ ($T$ = 760 K) ~ 2.5×10$^4$ ps]. They are sliced from a 3D model, with slice thickness 10 Å. One can see that although the labeled displacements are large, the IS $D_{min}$ is relatively small. This suggests that although the motions of individual atoms are significant, the global configuration is only slightly varied.

Figure 1(c) reinforces this observation by showing their distribution plot. It indicates that lots of atomic motions are position-replacing as suggested by the discrete peaks around 2.8 (the second peak) and 5 Å (the third) in the labeled displacements. As discussed previously (e..g, ref. [13]) that the peak and valleys in the displacement distributions $p(u)$ suggest that the atomic movements proceed in a cooperative manner, that is, one atom jumps to the position that is previously taken by another atom, such as its nearest or secondary neighbors. These can be seen from the fact that the peak positions of $p(u)$ match these of $g(r)$, the radial distribution function.

**Scaling relation between $\delta$ and IS $D_{min}$ in Al$_{85}$Sm$_{15}$.** To substantiate this proposal, we use the molecular dynamics simulations of dynamic mechanical spectroscopy (MD-DMS, see *Method* and Fig.S1) [13, 34, 35] to study the relaxation behaviors in a set of model glass forming liquids. We start with a model system that has a composition of Al$_{85}$Sm$_{15}$ based on the force-field of ref.[36]. In a previous work, Sun *et al*[37] have illustrated that this force-field can yield relaxation processes that are consistent with the experimental DMS results and predict an additional process.

Meanwhile, the MD-DMS method uses the same protocol as in DMS experiments, and it has been applied in studying several relaxation processes in glasses[13, 15, 34, 35, 37-39]. Thus, our simulations can be considered as an *in-silico* version of DMS experiments.

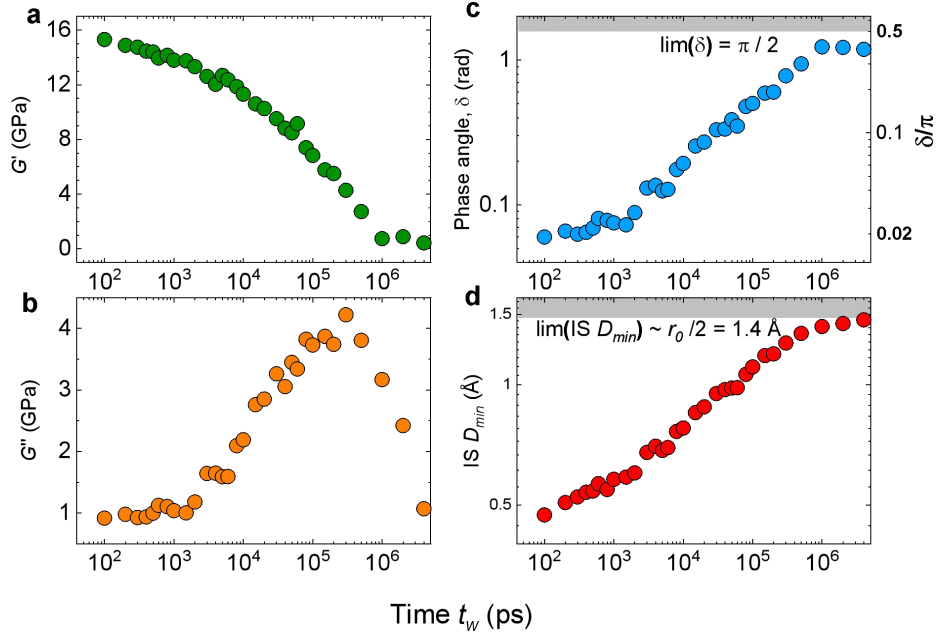

**FIG.2. Relaxation properties (a)** storage shear modulus $G'$, **(b)** loss shear modulus $G''$, **(c)** phase angle $\delta$, **(d)** IS $D_{min}$ in an Al$_{85}$Sm$_{15}$ metallic glass forming liquids at $T$ =740 K, and $p = 0$.

As a typical example, Fig. 2 presents the relaxation properties as a function of period time $t_w$ of the perturbation (which is inverse to the frequency $f = 1/t_w$) for temperature $T$ = 740 K and pressure $p$ = 0. It includes the storage shear modulus $G'$ in Fig.2(a), loss shear modulus $G''$ in Fig.2(b), phase angle $\delta$ in Fig.2(c) between the applied strain and response stress, as well as the IS $D_{min}$ in Fig.2(d). The decay of $G'$ [Fig.2(a)] and the broad peak of $G''$ [Fig.2(b)] are due to the so-called primary $\alpha$ relaxation. It is the dynamic process of the glass transition. For smaller $t_w \ll \tau_\alpha$ ($T$ = 740 K) = $3\times10^5$ ps (the peak time of $G''$) the system behaves solid-like having a small value of $\delta \to 0$, while the system is liquid-like ($\delta \to \pi/2$) for larger $t_w \gg \tau_\alpha$.

Both $\delta$ and IS $D_{min}$ have limiting value scopes: $\delta \in (0, \pi/2)$ where the low and high limited values correspond to the Hook elasticity ($\sigma \propto \gamma$) and Newtonian fluid ($\sigma \propto d\gamma/dt$) respectively; IS $D_{min} \in (0, \sim r_0/2)$ where $r_0$ is the average distance between the atoms for systems with isotropic interactions (e.g., without bonding).

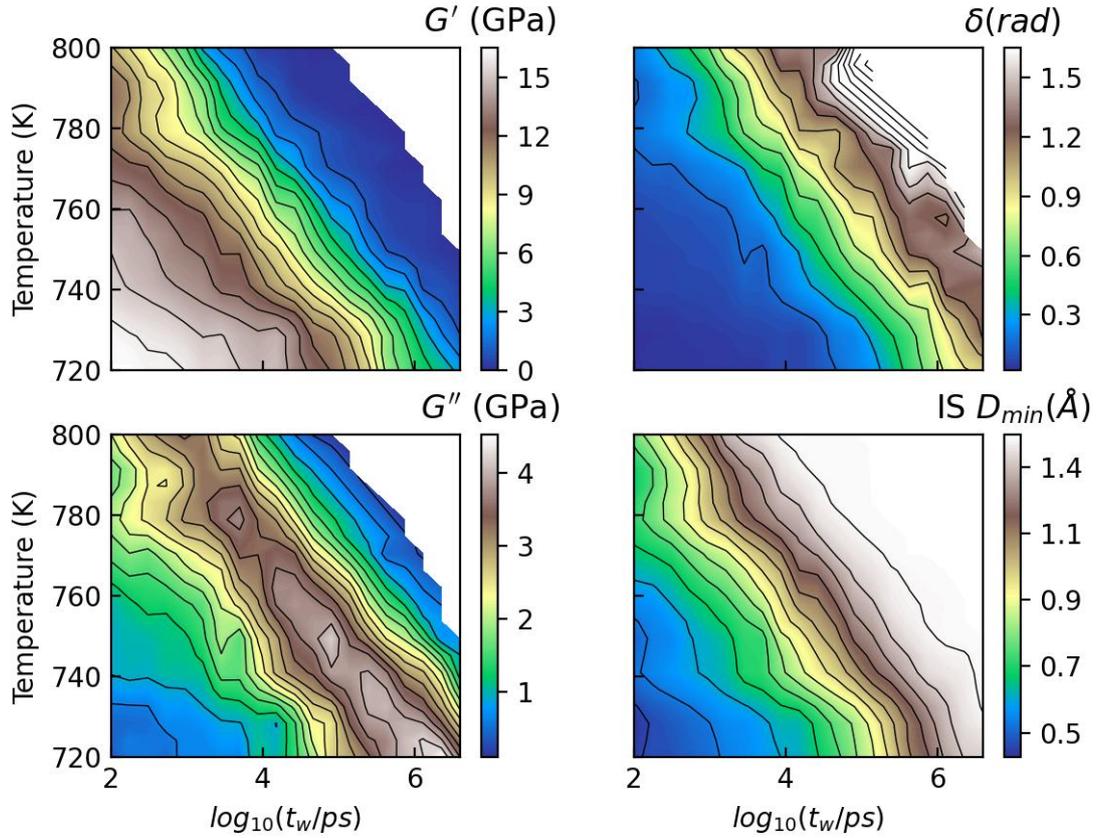

**FIG.3. Relaxation maps over wide-range of temperature and timescales.** (a)

storage shear modulus $G'$, **(b)** loss shear modulus $G''$, **(c)** phase angle $\delta$, **(d)** IS $D_{min}$ as two-dimensional functions of temperature and periodic time ($t_w$ in log-scale).

Figure 3 shows 2-dimensional contour plots of the relaxation properties over the temperature range from $T$ = 720 to 800 K at $p$ = 0. All the data are from the equilibrium supercooled liquid states (Fig.S2 and Fig.S3). We have examined a wide range of $t_w$, covering 5 orders of magnitude in timescales, from 100 ps to $2 \times 10^6$ ps (i.e., 2 microseconds). We did not include the short-time data ($t_w$ < 100ps), as they might be influenced by the contributions from atomic vibrations.

We note that the wide time-scale is essential for relating the simulations results to the experiments, as demonstrated in recent works[13, 37].

One can see from Fig.3(a) and (b) that the α relaxation moves to higher $t_w$ when the temperature is lowered, suggesting the sluggish dynamics in the supercooled liquids. These features mimic the experimental observations. It is interesting to find that the phase angle $\delta$ in Fig.3(c) and the IS $D_{min}$ in Fig.3 (d) show a similar dependence on temperature $T$ and period $t_w$.

Figure 4(a) plots the phase angle $\delta$ as a function of the IS $D_{min}$. It reveals a clear correlation between the two and all the data collapse on a single master line. A least square fitting to data yields a power law relation:

$$\delta = a \, [IS \, D_{min}]^b = A \, [IS \, D_{min} / r_0]^b \quad Eq.(1)$$

where the constants $a$, $A$ and $b$ are fitting parameters and the power $b$ = 2.8 ± 0.1. In addition, $r_0 = (V/N)^{1/3}$ is a typical length scale ($V$ and $N$ are the volume and number of particles respectively), it is introduced such that the parameter $A$ is dimensionless. This relation holds individually for each temperature as well (Fig.S4 and S5). In this work, we focus on the scaling behavior and the $b$ values, and do not delve into the details of $A$. As will be discussed later, this relationship could be explained based on PEL.

In additional, we note that some other functions such as the labeled (traditional) RMSD and the IS RMSD (i.e., the RMSD based on the inherent structures) are not correlated with phase angle (Fig. S6). Computing the direct $D_{min}$ (for instantaneous configurations, without evoking the minimization to obtain IS), we find that the direct $D_{min}$ and phase angle are also correlated, while the curve is nonlinear in the log-log plot (Fig.S6). In a previous work[35], we found that in the low temperature glassy state (non-equilibrium) δ is related to the number of atoms that jump with a distance larger than the half of the mean interatomic distance. Such a relation is not found to collapse the data for the supercooled liquid studied in this work (Fig.S7). This might due to that the atomic motions in supercooled liquids are more abundant and cooperative.

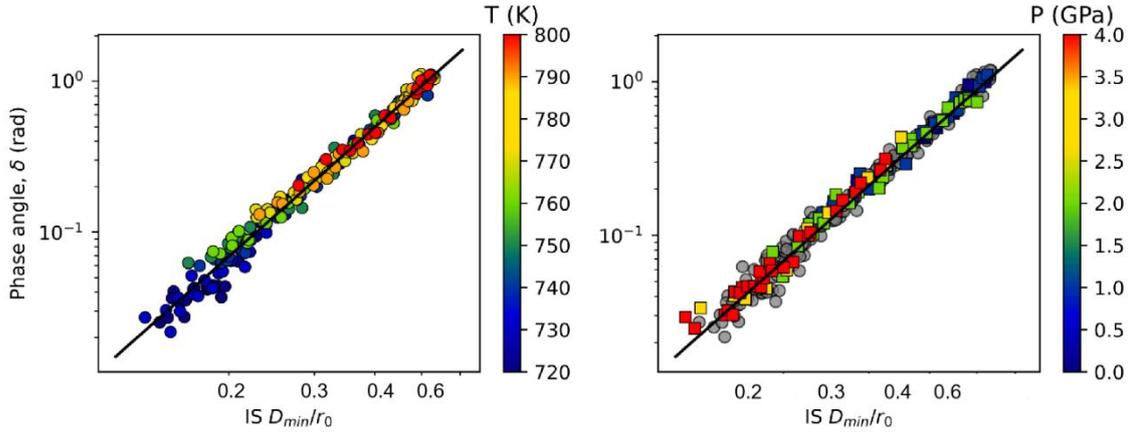

**FIG4. Scaling relation between $\delta$ and IS $D_{min}$.** **(a)** data for configurations at different temperatures and the pressure is fixed at $p = 0$; **(b)** data for configurations at different $p$ at a constant temperature $T = 800$ K as indicated by the color bar; the gray disk symbols are the same data from (a) $p = 0$.

These results suggest a fundamental relation between the relaxation property and the feature of PEL. It indicates that the phase angle is predominantly controlled by the configurations overlapping. On the other hand, the atomic motions as probed by the labeled RMSD contribute to the thermal fluctuation and diffusion of the system. But only those atomic motions that influence the IS could contribute to the experiment-probed phase angle.

**Generalization to different glass-forming liquids.** Besides varying temperature at a constant external pressure $p = 0$, we have changed thermodynamic states of the system (in the equilibrium supercooled liquids as well) by tuning the external pressure *p*. As a typical example, Fig.4 (b) shows the results that *p* in the range of [0, 4] GPa, while keeping the temperature a constant $T$ = 800 K. As can be seen in Fig.4(b), all the data points with different pressures (as indicated by color squares) are overlapped with those by changing the temperature [as indicated by gray circle symbols, which are the same in Fig.4(a)]. The relationship between IS $D_{min}$ and δ, i.e., Eq.1, fits these data within the same degree of accuracy. Other combinations of temperature and pressure give the same results.

We note that our findings are not limited to the $Al_{85}Sm_{15}$ model liquids; we have validated our findings in other six different model glass forming liquids, including three metallic models $Ni_{80}P_{20}$, $Pd_{80}Si_{20}$ and $Y_{65}Cu_{35}$ based on many-body embody atom method potential[40, 41], two models based on Lennard-Jones force-field, which are the Kob-Andersen (KA) model[42] and the Weeks-Chandler-Andersen (WCA) model[43], and a silica ($SiO_2$) model based on the three-body Tersoff force-field[44]. Results for these six systems are shown in Fig. 5. The detailed data are presented in Fig.S8-S13 and supplementary data. They confirm Eq.1 with reasonable accuracy. There are slight variations of the fitting parameters *A* and *b*, which might origin from the different interaction force-fields and the organizations of PEL.

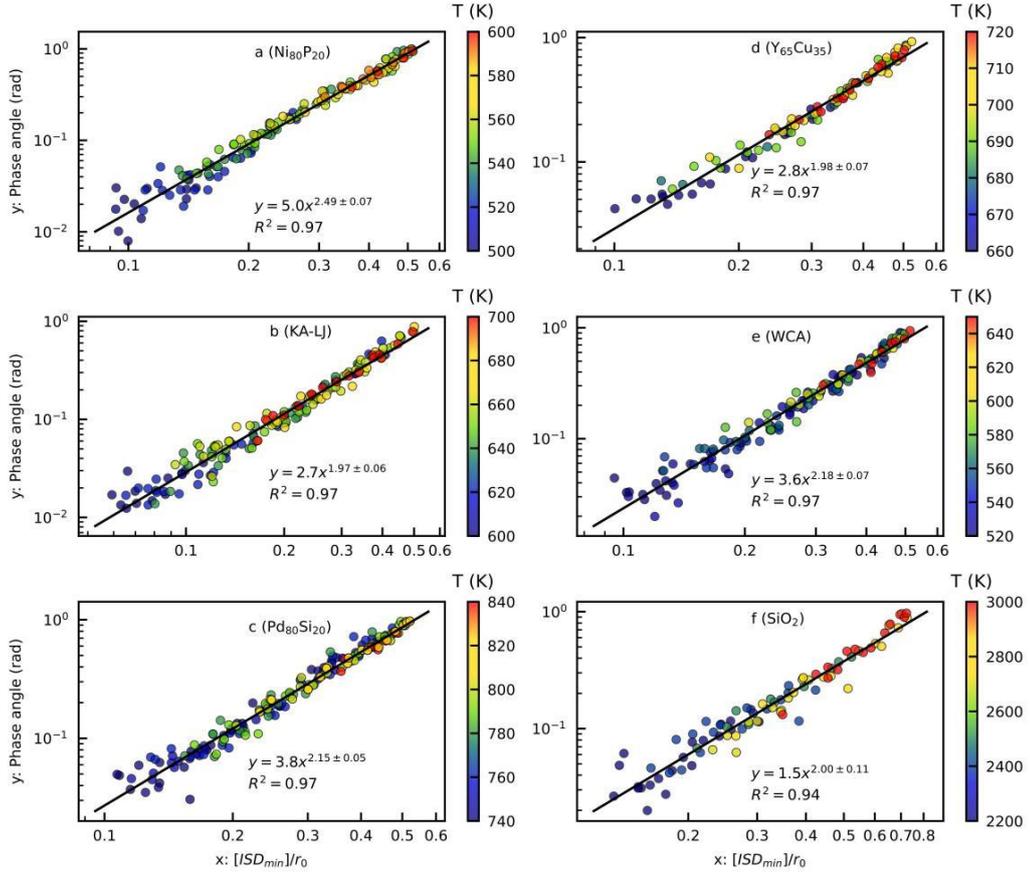

**FIG.5.** Relationship between $\delta$ and IS $D_{min}$ for other six different model glass forming liquids: (a) $Ni_{80}P_{20}$, (b) Kob-Andersen (KA) model, (c) $Pd_{80}Si_{20}$ (d) $Y_{65}Cu_{35}$ (e) Weeks-Chandler-Andersen (WCA) model and (f) $SiO_2$. For $SiO_2$ the upper limit of IS $D_{min}$ is larger than $r_0/2$ due to covalent bonding.

**Physical origin for the power-law scaling.** Now we are in the position to elucidate what the fundamental cause is for the power-law scaling between $\delta$ and IS $D_{min}$. One usually evoked approach is to study how the exponent changes with the physical dimension. Among our studied examples, the KA-LJ system can form glasses for both 2 and 3 dimensions, and we conducted additional computations for this purpose (Fig.S13). Interestingly, we find that $b$ ~2.0 is nearly the same for the 2 and 3 dimensional glasses (Fig.S13). This implies that the presented scaling is invariant over different dimensions, calling for a fundamental understanding.

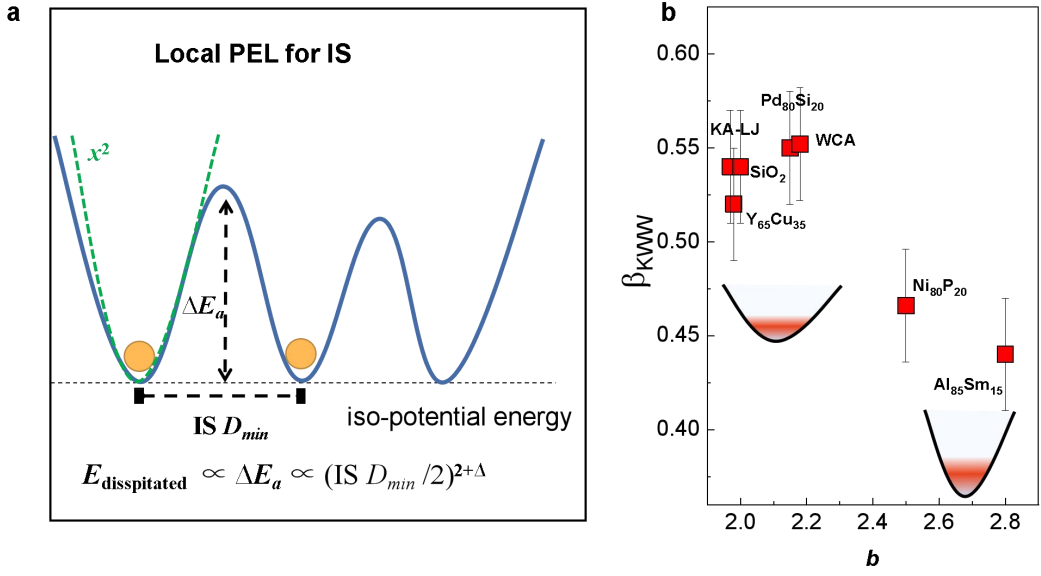

**FIG. 6. PEL interpretation for the scaling relation. (a)** Schematic PEL for dissipation relaxation. Relaxation damping is considered as a thermal activated process from two local PEL basins, which are separated by an energy barrier with the activation energy $\Delta E_a$. The green dashed line is a first approximation of the local PEL with a squared formula (i.e., the harmonic approximation). **(b)** Relation between $\beta_{KWW}$ and $b$ for the studied systems; the insets are schematic illustration for the local PEL for systems with lower and higher $b$ respectively.

One possible explanation might be given based on the feature of PEL. As schematically shown in Fig.6, IS $D_{min}$ characterizes the shortest distance between two ISs which are at the same potential level for equilibrium state as investigated in this work. Since dissipation damping is a thermodynamic activated process with an activation energy $\Delta E_a$. We might have $\Delta E_a \propto (IS\ D_{min})^{2+\Delta}$ where $\Delta$ is a number that describes the deviation of the PEL from the purely harmonic case. This is our theoretical premise based on PEL. Thus, we get $\delta \propto (IS\ D_{min})^{2+\Delta}$ and $b = 2+\Delta$ in Eq.1. Consequently, the value of $b$ in Eq.1 characterizes the local curvature of the inherent state in the PEL. The observed power-law scaling is now explained.

We fit the $G''(t_w)$ data with the Kohlrausch-Williams-Watts equation (KWW) $f(t)$

$\propto \exp[(-t/\tau)^{\beta_{KWW}}]$ and obtain the stretching parameter $\beta_{KWW}$ (Fig.S15)[45]. In the coupling model of Ngai, $n_{CM}$ = 1- $\beta_{KWW}$ describes the coupling degree for the relaxation dynamics[46]. It is interesting to find that Fig. 6(b) indicates a modest correlation between $\beta_{KWW}$ and $b$. Specifically, it shows that $Al_{85}Sm_{15}$ and $Ni_{80}P_{20}$ models exhibit large $b$ values and small $\beta_{KWW}$. This might reflect the local feature PEL in these systems, as shown in the insets of Fig.6(b). The results in Fig.6(b) are consistent with the coupling model[46].

**Discussion**

We have shown that by introducing a parameter IS $D_{min}$, which characterizes the variability of inherent structure in the PEL, one can interpret the mechanical phase angle $\delta$ in a simple-yet-quantitative scaling-law. Overall, these above results suggest a unified picture about the origin of mechanical phase angle $\delta$: no matter what the thermodynamic state of the system is, the value of $\delta$ is uniquely determined by the variability of inherent structures over the probing time $t_w$, as quantified by IS $D_{min}$. For example, at low temperature and shorter $t_w$, the value of $\delta$ is small as the variation between inherent structures is also small. The amorphous structures (packing, short- and medium-range orders) would influence $\delta$ through the change the population of unique inherent structures.

Since $\delta$ is central to DMS and relaxation dynamics, our results provide a fundamental basis for the application of DMS in relaxation dynamics investigations in glass-forming liquids. The emerging physical picture is that only the variability of inherent structure in PEL provides relaxation properties that can be probed by mechanical spectroscopy. Moreover, the IS $D_{min}$, as a newcomer to the theoretical toolbox, might also be utilized to the investigation of other issues (e.g., aging, crystallization, deformation and glass forming ability) in complex systems[47-50].

Finally, we anticipate that our discoveries could be investigated through experimental means, such as colloid experiments that can monitor particle positions, or by examining the contrast between relaxations and diffusions, which are primarily influenced by IS $D_{min}$ and RMSD, respectively.

**Methods**

**Simulations of model glass-forming liquids.**

An open-source molecular dynamics (MD) simulation code, LAMMPS, was used for all the simulations. As listed in Table I, we have studied 7 different glass-forming systems. All these systems contain $N = 9,088$ atoms. We used a *NPT* ensemble (that is, constant number of atoms, pressure, and temperature) for the for the equilibration of the studied configurations at different temperatures ($T$) and pressures ($p$). Particular attention has been paid that the simulations are long enough to ensure the configuration reach the desired states. Periodic boundary conditions were applied for all the simulations.

The inherent structures (ISs) are used in this work for the calculation of IS $D_{min}$. They are obtained by performing an energy minimization of the system, by iteratively adjusting atom coordinates. At that point the configuration will be in local potential energy minimum and the influence from atomic vibrations are removed. Specifically, we used the conjugate gradient (CG) algorithm as implemented in LAMMPS for the calculations. We have checked that other algorithm, such as 'fire' and 'quickmin' give the same results.

**TABLE 1.** Models studied in this work. These include metallic glass-forming models based on the embedded atom method (EAM) potential, Kob-Andersen model based on Lennard-Jones potential (KA-LJ), Weeks-Chandler-Andersen model based on Lennard-Jones potential (WCA-LJ) and the three-body tersoff potential for $SiO_2$.

| # | System | Force field | Conditions |
|---|--------|-------------|------------|
| 1 | $Al_{85}Sm_{15}$ | EAM | NPT($p=0$, and different $p$) |
| 2 | $Ni_{80}P_{20}$ | EAM | NPT ($p=0$) |
| 3 | $Y_{65}Cu_{35}$ | EAM | NPT ($p=0$) |
| 4 | $Pd_{80}Si_{20}$ | EAM | NPT ($p=0$) |

| 5 | $A_{80}B_{20}$ | KA-LJ | NVT ($N/V$=1.558) |
| 6 | $A_{80}B_{20}$ | WCA-LJ | NVT ($N/V$=1.558) |
| 7 | $SiO_2$ | tersoff | NPT ($p$=0) |

**Computation details about IS $D_{min}$.**

Briefly, we define a cost matrix for a pair of atoms $i$ and $j$, via Eq.2:

$$C_{i,j}(t) = | \mathbf{r_i}(t+t_0) - \mathbf{r_j}(t_0) |^2 \quad \text{Eq.2}$$

Let $X$ be a boolean matrix where $X[i, j] = 1$ if row $i$ is assigned to column $j$ and otherwise $X[i, j] = 0$. Then the IS $D_{min}$ can be calculated by optimizing the matrix $X$ for the minimization of the sum of the product of $C$ and $X$, as shown below in Eq. 3.

$$\text{IS } D_{min} = \left[\frac{1}{N} min \sum_{i,j}^{N,N} C_{i,j(t)} X_{i,j(t)}\right]^{1/2} \quad \text{Eq. 3}$$

In implementing the calculation of Eq. 3, we used the *linear_sum_assignment* function from the *scipy* package for python. It takes the cost matrix $C_{i,j}$ as one of the input parameters and returns the optimum assignments that minimize the sum of the product of $C$ and $X$ in Eq.3 (i.e., one does not need to construct $X$ by hand, more details can be found in the code). Obviously, if the optimum $X$ is the unit matrix, then Eq.3 is reduced to the common RMSD. This condition occurs when all atoms have small displacements ($\Delta r_i \ll r_0/2$). We note that the memory cost for the computation of IS $D_{min}$ is at the order $O(N^2)$.

**Dynamical mechanical spectroscopy and dissipation loss.**

The MD simulation of DMS (MD-DMS) was performed d on equilibrated configurations, covering a wide temperature range for the supercooled liquid. Specifically, at each configuration, we applied a sinusoidal shear strain

$$\gamma(t) = \gamma_A \sin(2\pi t/t_w) \quad \text{Eq. 4}$$

with an oscillation period $t_w$ (related to frequency $f = 1/t_w$) and a strain amplitude $\gamma_A$, along the *x-y* direction of the metallic glass forming liquid model. The resulting shear stress $\sigma(t)$ was computed and fitted according to

$$\sigma(t) = \sigma_0 + \sigma_A \sin(2\pi t/t_w + \delta) \qquad \text{Eq. 5}$$

Wherein, $\delta$ is the phase angle between stress and strain. The $\sigma_0$ is a linear term and is usually small. One typical example of the responding-stress is shown in Fig.S1, together with the fitted line by Eq. 5.

The strain was applied by a smooth change of the box shape, whereas stresses were directly measured from each component of the pressure. From these values, the storage ($G'$) and loss ($G''$) moduli are calculated as below

$$G' = \frac{\sigma_A}{\gamma_A}\cos(\delta); \; G'' = \frac{\sigma_A}{\gamma_A}\sin(\delta) \qquad \text{Eq. 6}$$

A strain amplitude $\gamma_A = 1.2\%$ was applied in all the MD-DMS simulations. Such a value of strain amplitude ensures the deformations do not change the structure and internal energy of the models. In other words, the deformation is in the linear response regime and the dissipation-fluctuation theorem is valid.

Technically, the *NVT* ensemble (constant number of atoms, volume, and temperature) was used during the MD-DMS simulations. Ten cycles were applied for each MD-DMS to measure the storage and loss shear moduli as well as damping factor δ, fitted by Eq. 6.

**Acknowledgments:** This work was supported by the National Science Foundation of China (NSFC 52071147) and the National Thousand Young Talents Program of China. K.S. received support from DFG via the Leibniz Program and SFB 1073. The computational work was carried out on the public computing service platform provided by the Network and Computing Center of HUST. We appreciate Prof. Ji-Chao Qiao (Northwestern Polytechnical University of China), Prof. Yang Sun (Xiamen University of China), Prof. Yu-Jiang Wang (Institute of Mechanics, Chinese Academy of Sciences) and Prof. Peng Tan (Fuda University of China) for advice, helps and discussions.

**Authors contributions**

H.-B.Y and K.S. proposed the research; H.-B.Y conducted the calculations and analyzed the data with the inputs from L.G. and J.-Q. G. The manuscript was jointly written by H.-B. Y. and K.S.

**Competing interests**

Authors have no competing interests.

**Data and materials availability**

All data is available in the manuscript or the supplementary information.

**Code availability:** A python script for the compunction of IS $D_{min}$ is available in the *supporting information* or at https://note.youdao.com/s/2RMcW3Oz

**References**

[1] H. Tanaka, H. Tong, R. Shi, J. Russo, Revealing key structural features hidden in liquids and glasses, Nat. Rev. Phys. **1**, 333–348 (2019).

[2] V. Bapst, T. Keck, A. Grabska-Barwińska, C. Donner, E.D. Cubuk, S.S. Schoenholz, A. Obika, A.W.R. Nelson, T. Back, D. Hassabis, P. Kohli, Unveiling the predictive power of static structure in glassy systems, Nat. Phys. **16**, 448–454 (2020).

[3] V.N. Manoharan, Colloidal matter: Packing, geometry, and entropy, Science **349**, 1253751 (2015).

[4] W. Anderson P, Through the Glass Lightly, Science **267**, 1615–1616 (1995).

[5] L. Berthier, G. Biroli, Theoretical perspective on the glass transition and amorphous materials, Rev. Mod. Phys. **83**, 587–645 (2011).

[6] H.B. Yu, W.H. Wang, H.Y. Bai, K. Samwer, The β-relaxation in metallic glasses, Natl. Sci. Rev. **1**, 429–461 (2014).

[7] Z. Wang, W.-H. Wang, Flow units as dynamic defects in metallic glassy materials, Natl. Sci. Rev. **6**, 304-323 (2018).

[8] J.C. Qiao, Q. Wang, J.M. Pelletier, H. Kato, R. Casalini, D. Crespo, E. Pineda, Y. Yao, Y. Yang, Structural heterogeneities and mechanical behavior of amorphous alloys, Prog. Mater. Sci. **104**, 250–329 (2019).

[9] Y. Gao, B. Zhao, J.J. Vlassak, C. Schick, Nanocalorimetry: Door opened for in situ material characterization under extreme non-equilibrium conditions, Prog. Mater. Sci. **104**, 53–137 (2019).

[10] J. Krausser, K.H. Samwer, A. Zaccone, Interatomic repulsion softness directly controls the

fragility of supercooled metallic melts, Proc. Natl. Acad. Sci. (U.S.A.) **112**, 13762-13767 (2015).

[11] Y.H. Sun, A. Concustell, A.L. Greer, Thermomechanical processing of metallic glasses: extending the range of the glassy state, Nat. Rev. Mater. **1**, 16039-16039 (2016).

[12] B. Ruta, E. Pineda, Z. Evenson, Relaxation processes and physical aging in metallic glasses, J. Phys.: Condens. Matter **29**, 503002-503002 (2017).

[13] H.-B. Yu, R. Richert, K. Samwer, Structural Rearrangements Governing Johari-Goldstein Relaxations in Metallic Glass, Sci. Adv. **3**, e1701577 (2017).

[14] B. Guiselin, C. Scalliet, L. Berthier, Microscopic origin of excess wings in relaxation spectra of supercooled liquids, Nat. Phys. **18**, 468-472 (2022).

[15] B. Wang, L.J. Wang, B.S. Shang, X.Q. Gao, Y. Yang, H.Y. Bai, M.X. Pan, W.H. Wang, P.F. Guan, Revealing the ultra-low-temperature relaxation peak in a model metallic glass, Acta Mater. **195**, 611–620-611–620 (2020).

[16] C. Chang, H.P. Zhang, R. Zhao, F.C. Li, P. Luo, M.Z. Li, H.Y. Bai, Liquid-like atoms in dense-packed solid glasses, Nature Materials **21**, 1240 (2022).

[17] P. Lunkenheimer, A. Loidl, B. Riechers, A. Zaccone, K. Samwer, Thermal expansion and the glass transition, Nature Physics **19**, 694–699 (2023).

[18] T. Hecksher, D.H. Torchinsky, C. Klieber, J.A. Johnson, J.C. Dyre, K.A. Nelson, Toward broadband mechanical spectroscopy, Proc. Natl. Acad. Sci. (U.S.A.) **114**, 8710-8715 (2017).

[19] S. Song, F. Zhu, M. Chen, Universal scaling law of glass rheology, Nat. Mater. **21**, 404–409 (2022).

[20] E.D. Cubuk, R.J.S. Ivancic, S.S. Schoenholz, D.J. Strickland, A. Basu, Z.S. Davidson, J. Fontaine, J.L. Hor, Y.R. Huang, Y. Jiang, N.C. Keim, K.D. Koshigan, J.A. Lefever, T. Liu, X.G. Ma, D.J. Magagnosc, E. Morrow, C.P. Ortiz, J.M. Rieser, A. Shavit, T. Still, Y. Xu, Y. Zhang, K.N. Nordstrom, P.E. Arratia, R.W. Carpick, D.J. Durian, Z. Fakhraai, D.J. Jerolmack, D. Lee, J. Li, R. Riggleman, K.T. Turner, A.G. Yodh, D.S. Gianola, A.J. Liu, Structure-property relationships from universal signatures of plasticity in disordered solids, Science **358**, 1033-1037 (2017).

[21] B. Mei, Y. Zhou, K.S. Schweizer, Experimental test of a predicted dynamics structure-thermodynamics connection in molecularly complex glass-forming liquids, Proc. Natl. Acad. Sci. (U.S.A.) **118**, e2025341118 (2021).

[22] P.G. Debenedetti, F.H. Stillinger, Supercooled liquids and the glass transition, Nature **410**, 259-267

(2001).

[23] M.D. Ediger, M. Gruebele, V. Lubchenko, P.G. Wolynes, Glass Dynamics Deep in the Energy Landscape, J. Phys. Chem. B **125**, 9052–9068 (2021).

[24] Y. Fan, T. Iwashita, T. Egami, Energy landscape-driven non-equilibrium evolution of inherent structure in disordered material, Nat. Commun. **8**, 15417 (2017).

[25] Q. Liao, L. Berthier, Hierarchical Landscape of Hard Disk Glasses, Phys. Rev. X **9**, 011049 (2019).

[26] Y. Nishikawa, M. Ozawa, A. Ikeda, P. Chaudhuri, L. Berthier, Relaxation Dynamics in the Energy Landscape of Glass-Forming Liquids, Phys. Rev. X **12**, 021001 (2022).

[27] P. Charbonneau, J. Kurchan, G. Parisi, P. Urbani, F. Zamponi, Fractal free energy landscapes in structural glasses, Nat. Commun. **5**, 3725 (2014).

[28] J. Ding, L. Li, N. Wang, L. Tian, M. Asta, R.O. Ritchie, T. Egami, Universal nature of the saddle states of structural excitations in metallic glasses, Mater. Today Phys. **17**, 100359-100359 (2021).

[29] P. Cao, M.P. Short, S. Yip, Potential energy landscape activations governing plastic flows in glass rheology, Proc. Natl. Acad. Sci. (U.S.A.) **116**, 18790-18797 (2019).

[30] C. Liu, Y. Fan, Emergent Fractal Energy Landscape as the Origin of Stress-Accelerated Dynamics in Amorphous Solids, Phys. Rev. Lett. **127**, 215502 (2021).

[31] G. Parisi, Order Parameter for Spin-Glasses, Phys. Rev. Lett. **50**, 1946-1948 (1983).

[32] B. Guiselin, G. Tarjus, L. Berthier, Static self-induced heterogeneity in glass-forming liquids: Overlap as a microscope, The Journal of Chemical Physics **156**, 194503 (2022).

[33] D.F. Crouse, On implementing 2D rectangular assignment algorithms, IEEE Transactions on Aerospace and Electronic Systems **52**, 1679-1696 (2016).

[34] H.-B. Yu, M.-H. Yang, Y. Sun, F. Zhang, J.-B. Liu, C.Z. Wang, K.M. Ho, R. Richert, K. Samwer, Fundamental Link between beta Relaxation, Excess Wings, and Cage-Breaking in Metallic Glasses, Journal of Physical Chemistry Letters **9**, 5877-5883 (2018).

[35] H.-B. Yu, K. Samwer, Atomic mechanism of internal friction in a model metallic glass, Phys. Rev. B **90**, 144201 (2014).

[36] H. Song, M.I. Mendelev, Molecular Dynamics Study of Mechanism of Solid-Liquid Interface Migration and Defect Formation in Al3Sm Alloy, JOM **73**, 2312–2319-2312–2319 (2021).

[37] Y. Sun, S.-X. Peng, Q. Yang, F. Zhang, M.-H. Yang, C.-Z. Wang, K.-M. Ho, H.-B. Yu, Predicting

Complex Relaxation Processes in Metallic Glass, Phys. Rev. Lett. **123**, 105701 (2019).

[38] B. Shang, J. Rottler, P. Guan, J.-L. Barrat, Local versus Global Stretched Mechanical Response in a Supercooled Liquid near the Glass Transition, Phys. Rev. Lett. **122**, 105501 (2019).

[39] G.J. Lyu, J.C. Qiao, Y. Yao, Y.J. Wang, J. Morthomas, C. Fusco, D. Rodney, Microstructural effects on the dynamical relaxation of glasses and glass composites: A molecular dynamics study, Acta Mater. **220**, 117293 (2021).

[40] H.W. Sheng, W.K. Luo, F.M. Alamgir, J.M. Bai, E. Ma, Atomic packing and short-to-medium-range order in metallic glasses, Nature **439**, 419-425 (2006).

[41] Q. Wang, J.H. Li, J.B. Liu, B.X. Liu, Favored Composition Design and Atomic Structure Characterization for Ternary Al–Cu–Y Metallic Glasses via Proposed Interatomic Potential, J. Phys. Chem. B **118**, 4442-4449 (2014).

[42] W. Kob, H.C. Andersen, Testing mode-coupling theory for a supercooled binary Lennard-Jones mixture I: The van Hove correlation function, Phys. Rev. E **51**, 4626-4641 (1995).

[43] J.D. Weeks, D. Chandler, H.C. Andersen, Role of Repulsive Forces in Determining the Equilibrium Structure of Simple Liquids, J. Chem. Phys. **54**, 5237-5247 (1971).

[44] S. Munetoh, T. Motooka, K. Moriguchi, A. Shintani, Interatomic potential for Si–O systems using Tersoff parameterization, Comp. Mat. Sci. **39**, 334-339 (2007).

[45] R. Böhmer, K.L. Ngai, C.A. Angell, D.J. Plazek, Nonexponential relaxations in strong and fragile glass formers, The Journal of Chemical Physics **99**, 4201-4209 (1993).

[46] K.L. Ngai, An extended coupling model description of the evolution of dynamics with time in supercooled liquids and ionic conductors, Journal of Physics: Condensed Matter **15**, S1107 (2003).

[47] Y. Wu, D. Cao, Y. Yao, G. Zhang, J. Wang, L. Liu, F. Li, H. Fan, X. Liu, H. Wang, X. Wang, H. Zhu, S. Jiang, P. Kontis, D. Raabe, B. Gault, Z. Lu, Substantially enhanced plasticity of bulk metallic glasses by densifying local atomic packing, Nat. Commun. **12**, 6582 (2021).

[48] F. Spieckermann, D. Sopu, V. Soprunyuk, M.B. Kerber, J. Bednarcik, A. Schokel, A. Rezvan, S. Ketov, B. Sarac, E. Schafler, J. Eckert, Structure-dynamics relationships in cryogenically deformed bulk metallic glass, Nat. Commun. **13**, 127-127 (2022).

[49] S.A. Kube, S. Sohn, R. Ojeda-Mota, T. Evers, W. Polsky, N. Liu, K. Ryan, S. Rinehart, Y. Sun, J. Schroers, Compositional dependence of the fragility in metallic glass forming liquids, Nat. Commun. **13**, 3708 (2022).

[50] A. Das, P.M. Derlet, C. Liu, E.M. Dufresne, R. Maass, Stress breaks universal aging behavior in a metallic glass, Nat. Commun. **10**, 5006 (2019).